\documentclass[a4paper,twoside]{article}

\usepackage{epsfig}
\usepackage{subcaption}
\usepackage{calc}
\usepackage{amssymb}
\usepackage{amstext}
\usepackage{amsmath}
\usepackage{amsthm}
\usepackage{multicol}
\usepackage{pslatex}
\usepackage{apalike}
\usepackage{SCITEPRESS}     % Please add other packages that you may need BEFORE the SCITEPRESS.sty package.

\begin{document}

\title{Artificial Neural Networks Jamming on the Beat}

\author{\authorname{Alexey Tikhonov\sup{1} and Ivan P. Yamshchikov\sup{2}\orcidAuthor{0000-0003-3784-0671}}
\affiliation{\sup{1}Yandex, Berlin, Germany}
\affiliation{\sup{2}Higher School of Economics, St. Petersburg, Russia}
\email{altsoph@gmail.com, ivan@yamshchikov.info}
}


\abstract{This paper addresses the issue of long-scale correlations that is characteristic for symbolic music and is a challenge for modern generative algorithms. It suggests a very simple workaround for this challenge, namely, generation of a drum pattern that could be further used as a foundation for melody generation. The paper presents a large dataset of drum patterns alongside with corresponding melodies. It explores two possible methods for drum pattern generation. Exploring a latent space of drum patterns one could generate new drum patterns with a given music style. Finally, the paper demonstrates that a simple artificial neural network could be trained to generate melodies corresponding with these drum patters used as inputs. Resulting system could be used for end-to-end generation of symbolic music with song-like structure and higher long-scale correlations between the notes.}

\onecolumn \maketitle \normalsize \setcounter{footnote}{0} \vfill

\section{\uppercase{Introduction}}
\label{sec:introduction}

In recent years, there have been many projects dedicated to neural network-generated music. For an extended survey of such methods see \cite{briot2019deep}. However there were several attempts to automate the process of music composition long before the era of artificial neural networks. The well-developed theory of music inspired many heuristic approaches to automated music composition. The earliest idea that we know of dates as far back as the nineteenth century, see \cite{lovelace}. In the middle of the twentieth century, a Markov chain approach for music composition was developed in \cite{hiller}, this approach became relatively popular and was revisited and improved in many followings works, see, for example, \cite{hill2011markov}. \cite{lin2017critical} have demonstrated that music, as well as some other types of human-generated discrete time series, tends to have long-distance dependencies that cannot be captured by models based on Markov-chains. Recurrent neural networks (RNNs) seem to be better at processing data series with longer internal dependencies \cite{sundermeyer}, such as sequences of notes in tune, see \cite{bl}. 

Indeed, a variety of different recurrent neural networks such as hierarchical RNN, gated RNN, Long-Short Term Memory (LSTM) network, Recurrent Highway Network, etc., were successfully used for music generation in \cite{chu}, \cite{colombo}, \cite{johnson}, \cite{wu2019hierarchical}, \cite{lattner2019high}. Google Magenta released a series of projects dedicated to music generation. In particular, one should mention a music\_vae model \cite{roberts2018hierarchical} that could be regarded as an extension of drum\_rnn\footnote{https://github.com/tensorflow/magenta}. It is important to distinguish the generative models like music\_vae and the generative models for music that use a straightforward language model approach and predict the next sound using the previous one as an input. For example, \cite{choi} used a language model approach to predict the next step in a beat with an LSTM. Variational autoencoder (VAE), see \cite{bowman} and \cite{semeniuta}, on the other hand, allows us to construct a latent space in which each point corresponds to a melody. Such spaces obtained with VAE or any other suitable architecture are of particular interest for different tasks connected with computational creativity since they can be used both to study and classify musical structures, as well as to generate new tunes with specified characteristics. 

Generation of polyphonic music is more challenging that generation of a single melody line. \cite{lyu2015modelling} uses a set of parallel, tied-weight recurrent networks designed to be invariant to transpositions. \cite{johnson2017generating} generate music in two steps. First, a chord LSTM predicts a chord progression based on a chord embedding. A second LSTM then generates polyphonic music from the predicted chord progression. \cite{chuan2018modeling} present an approach for predictive music modeling and music generation that incorporates domain knowledge in its representation. Majority of the polyphonic methods that we know of still face certain difficulties with temporal structure of the generated tunes. These problems are fundamentally connected with properties of recurrent networks when applied in an end-to-end setting. 

In this paper, we propose a straight-forward approach the addresses temporal challenges and could be used for end-to-end generation of tracks with song structure. We construct a latent explorable drum pattern space with some recognizable genre areas. We test two different smoothing methods are used on the latent space of representations. The obtained latent space is then used to sample new drum patterns. We experiment with two techniques, namely, variational autoencoder and adversarially constrained autoencoder interpolations (ACAI) \cite{berthelot2018understanding}. We also train a system that uses drum files as inputs and generates melodies corresponding to the input drum patterns. Such loops of melodies and drums could be combined into longer song-like tracks in the given tempo and tonality.

The contribution of this paper is three-fold: (1) we publish a large dataset of drum patterns, (2) develop an overall representation of typical beat patterns mapped into a two-dimensional space, and (3) demonstrate that a simple artificial neural network could be trained to generate tunes corresponding to the drum patterns.

\section{\uppercase{Dataset}}
\label{sec:1}
Most of the projects that we know of used small datasets of manually selected and cleaned beat patterns. One should mention a GrooveMonkee free loop pack\footnote{https://groovemonkee.com/collections/midi-loops}, free drum loops collection\footnote{https://www.flstudiomusic.com/2015/02/35-free-drum-loops-wav-midi-for-hip-hop-pop-and-rock.html} and aq-Hip-Hop-Beats-60–110-bpm\footnote{https://codepen.io/teropa/details/JLjXGK} or \cite{gillick2019learning}.

Unfortunately, majority of these datasets are either restricted to one or two specific genres or contain very limited amount of midi samples that does not exceed a dozen per genre. This amount of data is not enough to infer a genre-related latent space. Inferring this space, however, could be of utmost importance.  Due to the interpolative properties of the model that could work on such space, one can produce infinitely diverse patterns that still adhere to the genre-specific macro-structure. Groove MIDI \cite{gillick2019learning} to a certain extent goes in line with the material presented in the papers yet it is not big enough for the inference of the genre.

Here we introduce a completely new dataset of MIDI drum patterns\footnote{https://github.com/altsoph/drum\_space} that we automatically extracted from a vast MIDI collection available online. This dataset is based on approximately two hundred thousand MIDI files, and as we show later is big enough to infer the macroscopic structure of the underlying latent space with unsupervised methods. 

\subsection{Data filtering}

The pre-processing of the data was done as follows. Since the ninth channel is associated with percussion according to the MIDI standard, we assumed that we are only interested in the tracks that have non-trivial information in it. All the tracks with trivial ninth channels were filtered out. This filtering left us with almost ninety thousand tracks. Additional filtering included an application of a 4/4 time signature and quantization of the tracks. We are aware that such pre-processing is coarse since it ultimately corrupts several relatively popular rhythmic structures, for example, waltzes, yet the vast majority of the rhythmic patterns are still non-trivial after such pre-processing. We believe that  4/4 time signature is not a prerequisite for the reproduction of the results demonstrated here and encourage researchers to experiment and publish broad and diverse datasets of percussion patterns. In order to reduce the dimensionality of the problem, we have simplified the subset of instruments merging the signals from similar instruments. For example, all snares are merged into one snare sound, low and low mid- toms into a low tom, whereas and high tom and high mid-tom into a high tom. Finally, we had split the percussion tracks into percussion patterns. Every track was split into separate chunks based on long pauses. If a percussion pattern that was thirty-two steps long occurred at least three times in a row, it was added to the list of viable patterns. Trivial patterns with entropy below a certain minimal threshold were discarded from the list of viable patters. Finally, every pattern was checked to be unique in all its possible phase shifts. The resulting dataset includes thirty-three thousand of unique patterns in the collection and is published alongside this paper which is an order of magnitude larger that midi available data sources.

\begin{table}
\centering
\small{\begin{tabular}{l}
 \hline
 \\
// Filtering original MIDI dataset\\
\textbf{for}~~\verb"new_track"~~\textbf{in}~~\verb"MIDI_dataset"~~\textbf{do}\\
\textbf{if}~~\verb"new_track[channel_9] is non-trivial"\\
~~// Quantize with 4/4 signature\\
~~~~\verb"drum_track" $\leftarrow$ \verb"new_track[channel_9].quantize()"\\
~~// Merge different drums according to a predefined table\\
~~~~\verb"drum_track.merge_drums()"\\
~~// Split drum track into chunks\\
~~~~\textbf{for}~\verb"new_chunk"~\textbf{in}~\verb"drum_track.split_by_pauses()"~\textbf{do}\\
~~~~~~\textbf{if}~~\verb"length(new_chunk) == 32  \"\\
~~~~~~~~~~~~~~\textbf{and}~~\verb"new_chunk"$^3$ $\in$ \verb"drum_track  \"\\
~~~~~~~~~~~~~~\textbf{and}~~\verb"entropy(new_chunk)>k"\\
~~~~~~~~~\verb"percussion_patterns.append(new_chunk)"\\
\\
// Filtering non-unique percussion patterns\\
\textbf{for}~~\verb"new_pattern"~~\textbf{in}~~\verb"percussion_patterns"~~\textbf{do}\\
~~// Create all possible shifts of a pattern\\
~~\verb"shifted_patterns" $\leftarrow$ \verb"new_pattern.all_shifts()"\\
~~//Search for patterns that duplicate and delete them\\
~~\textbf{for}~~\verb"pattern"~~\textbf{in}~~\verb"percussion_patterns"~~\textbf{do}\\
~~~~\textbf{if}~~\verb"pattern"~$\in$~\verb"shifted_patterns"\\
~~~~~~\textbf{delete}~~\verb"pattern"\\
~~\verb"[new_pattern] + percussion_patterns"\\
\\
 \hline
\end{tabular}}
\caption{Pseudo-code that describes filtering heuristics used to form the dataset of percussion patterns.}
  \label{tab:code}
\end{table}

\subsection{Data representation}
\label{sec:drumdata}

\begin{figure}
% Use the relevant command to insert your figure file.
% For example, with the graphicx package use
 \includegraphics[width=0.5\textwidth]{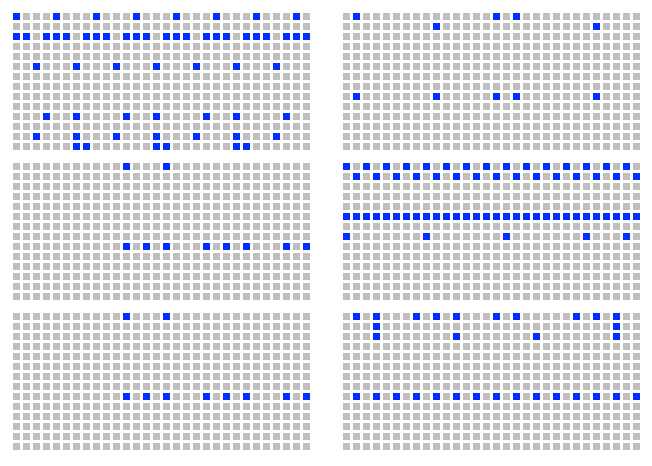}
% figure caption is below the figure
\caption{
Some examples of two-dimensional representation for drum patterns.}
\label{fig:1}    % Give a unique label
\end{figure}

The resulting dataset consists of similarly structured percussion patterns. Each pattern has thirty-two-time ticks for fourteen possible percussion instruments left after the simplification. Each pattern could be represented as a $14 \times 32$ matrix with ones on the positions, where corresponding instruments makes a hit. Figure \ref{fig:1} shows possible two-dimensional representations of the resulting patterns. 

We can also list all possible combinations of fourteen instruments that can play at the same time tick. In this representation, each pattern is described by thirty-two integers in the range from 0 to 16383. Such representation is straightforward and could be convenient for processing of the data with modern models used for generation of discrete sequences (think of a generative model with a vocabulary consisting of $2^{14}$ words). The dataset final dataset is published in the following format:
\begin{itemize}
\item the first column holds the pattern code that consists of thirty-two comma-separated integers in the range of $[0, 16383]$;
\item the second column holds four comma-separated float values that represent the point of this pattern in the latent four-dimensional space, that we describe below;
\item the third column holds two comma-separated float values of the t-SNE mapping from the four-dimensional latent space into a two dimensional one, see details below.
\end{itemize}

The model that we describe further works with a two-dimensional representation shown in Figure \ref{fig:1}.

\section{\uppercase{Models and experiments}}
\label{model}

In this papers we experiment with different autoencoders. Let us first briefly clarify the underlying principles of these architectures.

\subsection{Autoencoders}

Autoencoders are a broad class of structures that process the input $x \in \mathbb{R}^{d_x}$ through an 'encoder' $z = f_\theta (x)$ parametrized by $\theta$ to obtain a latent code $z \in \mathbb{R}^{d_z}$. The latent code is then passed through a 'decoder' $\hat{x}= g_\phi (z)$ parametrized by $\phi$ to produce an approximate reconstruction $\hat{x} \in \mathbb{R}^d_x$ of the input $x$. In this paper $f_\theta$ and $g_\phi$ are multi-layer neural networks. The encoder and decoder are trained simultaneously (i.e. with respect to $\theta$ and $\phi$) to minimize some notion of distance between the input $x$ and the output
$\hat{x}$, for example the squared L2 distance $||x - \hat{x} ||^2$.

Interpolating using an autoencoder describes the process of using the decoder $g_\phi$ to decode a mixture
of two latent codes. Typically, the latent codes are combined via a convex combination, so that
interpolation amounts to computing $\hat{x}_\alpha = g_\phi (\alpha z_1 +(1 - \alpha) z_2)$ for some $\alpha \in [0, 1]$ where $z_1 = f_\theta(x_1)$ and $z_2 = f_\theta(x_2)$ are the latent codes corresponding to data points $x_1$ and $x_2$. Ideally, adjusting $\alpha$ from $0$ to $1$ will produce a sequence of realistic datapoints where each subsequent $\hat{x}_\alpha$ is progressively less semantically similar to $x_1$ and more semantically similar to $x_2$. The notion of 'semantic similarity' is problem-dependent and ill-defined.

VAE assumes that the data is generated by a directed graphical model $p_\theta (x|h)$ and that the encoder is learning an approximation $q_\phi (h|x)$ to the posterior distribution $p_\theta (h|x)$. This yields an additional loss component and a specific training algorithm called Stochastic Gradient Variational Bayes (SGVB), see \cite{rezende} and \cite{kingma}. The probability distribution of the latent vector of a VAE typically matches that of the training data much closer than a standard autoencoder. 

ACAI has different underlying mechanism. It uses a critic network, as is done in Generative Adversarial Networks
(GANs) \cite{goodfellow2014generative}. The critic is fed interpolations of existing datapoints (i.e. $\hat{x}_\alpha$ as defined above). Its goal is to predict $\alpha$ from $\hat{x}_\alpha$. This could be regarded as a regularization procedure which encourages interpolated outputs to appear more realistic by fooling a critic network which has been trained to recover the mixing coefficient from interpolated data.

\subsection{Architecture}

\begin{figure*}
% Use the relevant command to insert your figure file.
% For example, with the graphicx package use
 \includegraphics[width=\textwidth]{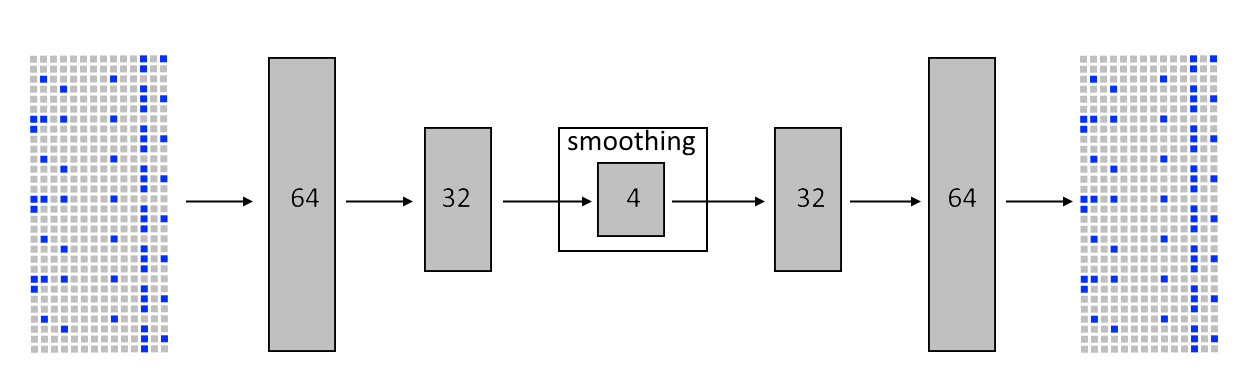}
% figure caption is below the figure
\caption{
Basic scheme of an autoencoder used to produce a latent space of patterns.}
\label{fig:2}    % Give a unique label
\end{figure*}

In this paper, we experiment with a network that consists of a 3-layered fully connected convolutional encoder, and a decoder of the same size. The encoder maps the beat matrix (32*14 bits) into four-dimensional latent space. The first hidden layer has sixty-four neurons; the second one has thirty-two. The ReLU activations are used between the layers, and a sigmoid maps the decoder output back into the bit mask. Figure \ref{fig:2} shows the general architecture of the network.

The crucial part of the model that is valid for further experiments is the space of latent codes or the so-called  'bottle-neck' of the architecture shown in Figure \ref{fig:2}. This is a four-dimensional space of latent representations $z \in \mathbb{R}^4$. The structural difference between the VAE and ACAI models with which we experiment further occurs exactly in this bottle-neck. The architectures of the encoder $f_\theta$ and decoder $g_\phi$ are equivalent. Effectively, VAE and ACAI could be regarded as two smoothing procedures over the space of latent codes.

\subsection{Vizualization of the obtained latent space}

\begin{figure}
% Use the relevant command to insert your figure file.
% For example, with the graphicx package use
 \includegraphics[width=0.5\textwidth]{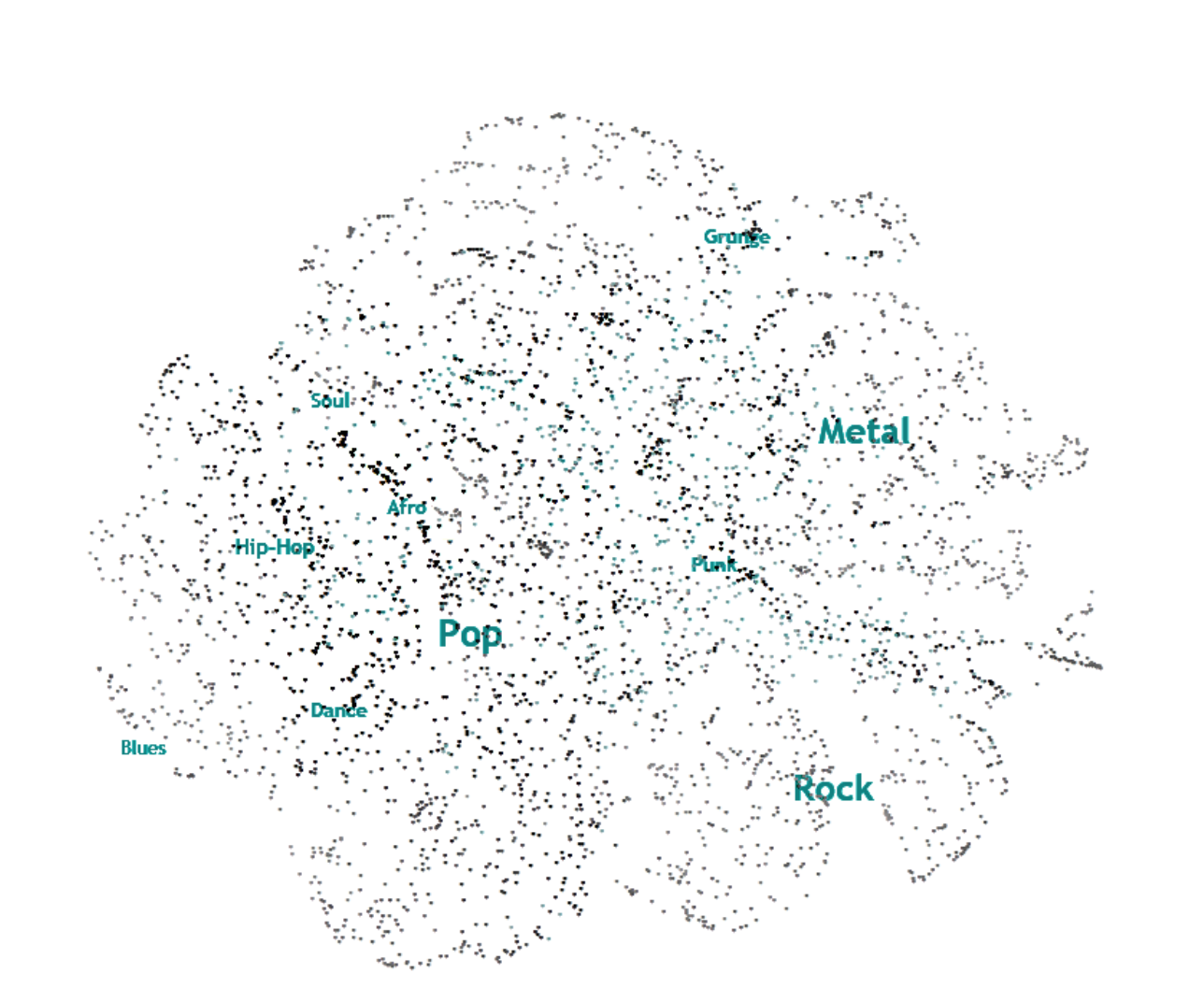}
% figure caption is below the Figure
\caption{
t-SNE projection of the latent percussion space produced by VAE. Different areas correspond to specific genres. One can see a clear macro-structure with hip-hop, soul an afro beats grouped closer together and with rock, punk and metal in another area of the obtained space.}
\label{fig:ill}    % Give a unique label
\end{figure}

To explore the obtained dataset, we have built an interactive visualization that is available online\footnote{http://altsoph.com/pp/dsp/map.html}.
and is similar to the one described in \cite{Ano18}. This visualization allows us to navigate the resulting latent space of percussion patterns. Training patterns are marked with grey and generated patterns are marked with red. For the interactive visualization, we use a t-SNA projection of the VAE space since it has a more distinctive geometric structure, shown in Figure \ref{fig:ill}.

Moreover, this visualization, in some sense, validates the data representation proposed above. Indeed, coarsely a third of tracks in the initially collected MIDIs had genre labels in filenames. After training VAE we used these labels to locate and mark the areas with patterns of specific genres. Closely looking at Figure \ref{fig:ill} that shows a t-SNE projection of the obtained latent space, one can notice that the geometric clusters in the obtained latent space correspond to the genres of the percussion patterns. The position of the genres on the Figure were determined by the mean of coordinated of the tracks attributed to the corresponding genre. One can see that related genres are closer to each other in the obtained latent space and the overall structure of the space is meaningful. For example the cloud of 'Punk' samples is located between 'Rock' and 'Metal' clouds, whereas 'Hip-Hop' is bordering 'Soul', 'Afro' and 'Pop'. The fact that VAE managed to capture this correspondence in an unsupervised set-up (as a by-product of training with a standard reconstruction loss) demonstrates that chosen data representation is applicable to the proposed task, and the proposed architecture manages to infer a valid latent space of patterns.

As we have mentioned above, we compare two different latent space smoothing techniques, namely, VAE and ACAI. It is important to note here that the standard VAE produces results that are good enough: the space mapping is clear and meaningful, as we have mentioned above. At the same time, the ACAI space seems to be smoother, yet harder to visualize in two dimensions. 

\begin{figure*}[h!]
 \includegraphics[width=\textwidth]{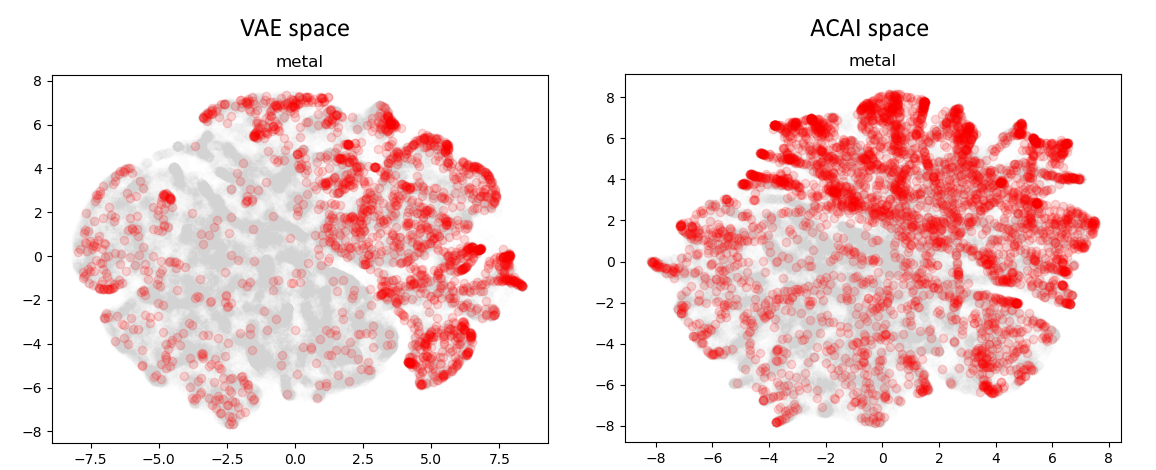}
% figure caption is below the figure
\caption{
The beats from the area that corresponds to the genre metal on the VAE space projection (left) and the ACAI space projection (right). VAE maps the tracks of the same genre closer together and therefore is beneficial for the visualization of the latent space.}
\label{fig:3}    % Give a unique label
\end{figure*}

Figure \ref{fig:3} illustrates this idea, showing the two-dimensional t-SNE mapping of the latent spaces produced by both methods with patterns that correspond to the genre METAL marked with red dots.  One can see that ACAI mapping of a particular genre is not as dense as VAE.  Due to this reason, we use t-SNE projection of VAE space for the interactive visualization mentioned above and throughout this paper.

However, we argue that the latent space produced with ACAI is better to sample from and discuss it in detail further. 

\subsection{Generating the beat}

The majority of the auto-encoder based methods generates new samples according to the standard logic. One can sample an arbitrary point from the latent space and use the decoder to convert that point into a new pattern. In the case of VAE one can also narrow the area of sampling and restrict the algorithm in the hope of obtaining beats that would be representative of the style typical for that area. However, an objective metric that could be used for quality estimation of the generated samples is still a matter of discussion. Such objective estimations are even harder in this particular case since the patterns are quantized and consist of thirty-two steps and fourteen instruments. Indeed, virtually any sequence could be a valid percussion pattern, and human evaluation of such tasks is usually costly and, naturally, subjective. We invite the reader to estimate the quality of the generated samples on her own using the demo mentioned above. At the same time we propose a simple heuristical method that allows putting the quality of different architectures into relative perspective.

Table \ref{tab:code} contains pseudo-code that was used for the filtering of the original MIDI dataset. We suggest using percussion related part of this filtering heuristic to estimate the quality of generated percussion patterns. Indeed one can generate a set of random points in the latent space, sample corresponding percussion patterns with the decoder, and then apply the filtering heuristics. The resulting percentage of the generated beats that pass the filter could be used as an estimate of the quality of the model.

The percentage of the real MIDI files from the training dataset that pass the final entropy filter could be used as a baseline for both architectures. 

To have a lower baseline, we also trained a classic auto-encoder without any smoothing of the latent space whatsoever. The examples of the tracks generated by it are also available online\footnote{https://github.com/altsoph/drum\_space}.

\begin{table}
\centering
\small{\begin{tabular}{ll}
 \hline
Model & \% of patterns after filtering\\
\\
AE & 28\%\\
VAE & 17\%\\
ACAI & 56\%\\
Empirical patterns & 82\%\\
 \hline
\end{tabular}}
\caption{Comparison of the two smoothing methods. ACAI seems to be way more useful for sampling since it produces a valid percussion pattern out of a random point in the latent space more than 50\% of the time and is three times more effective than VAE based architecture. In terms of the heuristic entropy filter, VAE performs even worse than AE, generating a lot of "dull" samples with entropy below the threshold.}
  \label{tab:dataform}
\end{table}

 This simple heuristic filtering shows that VAE-generated beats have a quality of about 17\%. In other words, on average, one out of six generated beats passes the simple filter successfully. In the case of ACAI, quality happens to be significantly higher. Namely, 56\% of the produced beats satisfy the filtering conditions. More than half of the generated patterns passed the filters.

In order to have a baseline to compare both methods, one can look at the percentage of empirical MIDI files that pass through the last entropy-based filter. One can see that in this context the patterns randomly sampled with ACAI are comparable with the empirical ones that were present in the original MIDI dataset.

\subsection{Jamming on the beat}

Filtering original data we could not help to notice that there are melodic patterns attached to every drum loop. These melody patterns were played by various instruments designated in a midi file. We collected the dataset of these melody loops and present them along with the drum patterns. Every melody is encoded with three different embeddings that we concatenate in the input vector. These embeddings include the instrument playing the melody, the tone and the octave. To simplify the generation we do not include the length of the note played, this could be easily amended with an additional embedding for the length pf a note. For a detailed description of the melody preprocessing we address the reader to \cite{yamshchikov2020music}.

\begin{figure}[h]
 \includegraphics[width=0.48\textwidth]{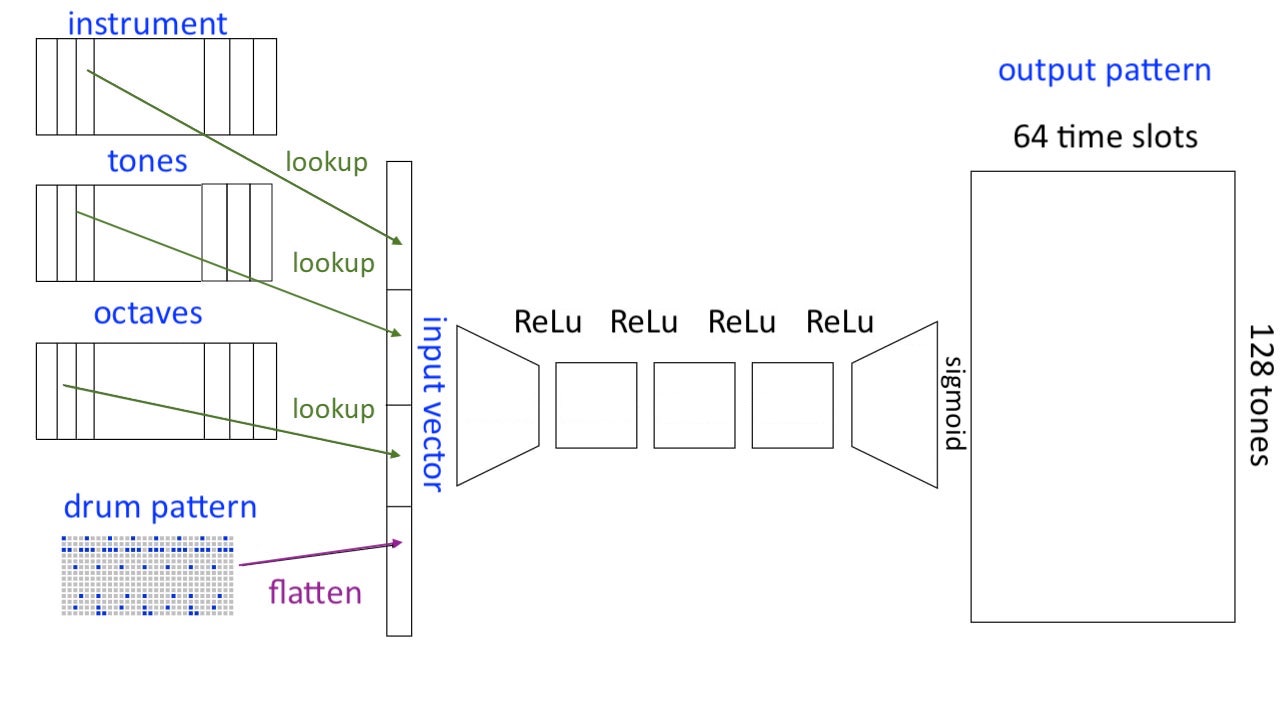}
% figure caption is below the figure
\caption{
A simple melody generation that generates melodies corresponding to the rhythm.}
\label{fig:4}    % Give a unique label
\end{figure}

These melody embedding are concatenated with the flattened representation of the beat and form an input vector, see Figure \ref{fig:4}.

Let us briefly describe Figure \ref{fig:4}. The input is represented as a 496 dimensional vector. It consist of a flattened drum pattern encoded as a $32 \times 14$ matrix, see Section \ref{sec:drumdata}, concatenated with three 16-dimensional embeddings of the instrument, tonality and octave. This input vector is first projected with $496 \times 64$ ReLu layer, that is processed through three more $64 \times 64$ ReLu layers, and a final $64 \times 8192$ sigmoid projection. The resulting 8192 vector is encoding 64 time steps with 128 possible pitches on every step.

As well as in other computational creativity tasks, see, for example, \cite{agafonova2020paranoid}, one  could significantly improve the results of the generation with automated heuristic filtering. In the case of symbolic music additional heuristic filtering of the generated melodies includes some basic rules that help to harmonize obtained tunes, i.e. if the tune dynamic exceeds the diapason of two octaves we filter it. Such cases are very rare for a given instrument within as relatively short drum loop. Another example of the heuristic could be: if the relative length of notes in a tune exceeds some fixed constant, say the shortest time between to notes is ten times shorter that the longest one, we could also filter that melody assuring a relatively dense tune on the beat without pauses lasting a half of the loop. These heuristics could be tweaked and tinkered with. The resulting melody patterns that we showcase in the repository are filtered with the following list of heuristics. We filter the loop out if:

\begin{itemize}
    \item it has sounds on three or less steps out of 64 time ticks;
    \item if there is a pause that is longer that 16 consecutive ticks;
    \item it includes two notes within a tone or a semitone from one another that are played at once;
    \item there are more than two octaves between the lowest and the highest note in a loop;
    \item the loop includes less than three different notes;
    \item at some point of the loop there are four or more notes played an once;
    \item the most frequent note in the loop makes up for more than three quarters of all notes played in the loop;
    \item the tonality detection heuristics detects a tonality that is different for the target tonality that was given in the input.
\end{itemize}

We developed a stand-alone tonality detector that is rather intuitive yet was not applied to the task of music generation before. To improve usability of the heuristics and reproducibility of the work that is especially important due to its explorative nature we publish our code online\footnote{https://github.com/altsoph/tonika\_detector}.
Tonality heuristics lists all possible tonalities. For every given melody we score the notes and consecutive note pairs that correspond to possible tonalities and choose the most probable one. You could test the heuristic on the dataset of midi files with labelled tonalities that is included in the repository.

The main idea behind melody generation that we wanted to illustrate here is that drum pattern could be an organizing structural layer used to achieve certain synchronisation on a longer time scale and therefor simplifying the generation of longer song-like structures, based on an interplay of certain repetitive patterns.

\section{Discussion}

Deep learning enables the rapid development of various generative algorithms. There are various limitations that hinder the arrival of algorithms that could generate discrete sequences that would be indistinguishable from the corresponding sequences generated by humans. In some contexts, the potential of such algorithms might still be limited with the availability of training data; in others, such as natural language, the internal structure of this data might be a challenge; finally, some of such tasks might be simply too intensive computationally and therefore too costly to use. However, percussion patterns do not have such limitations. The structure of the data can be formalized reasonably well and without significant loss of nuance. In this paper, we provide thirty-three thousand thirty-two step 4/4 signature percussion drums and demonstrate that such a dataset allows training a good generative model. We hope that as more and more data is available for experiments, percussion could be the first chapter to be closed in the book of generative music.

Once the drum pattern progression is defined one could use very intuitive generative methods in combination with tonality filter to generate the melody over the generated beat. The resulting loops could be rearranged in any progression providing longer macro structures. The level of novelty could be further controlled through purely combinatoric parameters of the song structure. These endeavours, however, are mostly either heuristically motivated or anecdotal rather than data-driven. Generative models capable of smooth interpolations between different rhythmic patterns represent another set of new research questions. Finally, nuances of percussion alongside with the datasets and the models that could capture these nuances, for example see \cite{gillick2019learning}, need further research. 

Aside from the generation of the symbolic music there are plenty of open questions regrading generation of sound fonts, intonation and dynamics both on micro and macro levels of the melody progression.

\section{Conclusion}

This paper presents a new huge dataset of MIDI percussion patterns that could be used for further research of generative percussion algorithms. Ir also presents corresponding melody patterns that are aligned with the given percussion tracks. This dataset could be further used for symbolic loop generation.

The paper also explores two autoencoder based architectures that could be successfully trained to generate new MIDI beats. Both structures have similar fully connected three-layer encoders and decoders but use different methods for smoothing of the produced space of latent representations. Adversarially constrained autoencoder interpolations (ACAI) seem to provide denser representations than the ones produced by a variational autoencoder. More than half of the percussion patterns generated with ACAI passes the simple heuristic filter used as a proxy for the resulting generation quality estimation. To our knowledge, this is the first application of ACAI to drum-pattern generation.

The interactive visualization of the latent space is available as a tool to subjectively assess the quality of the generated percussion patterns. 

Finally, the paper explores the possibility to generate melodies that correspond to the given input pattern and demonstrates that this could be done with a relatively straight-forward artificial neural network.

\section*{Acknowledgements}

The authors would like to thank Valentina Barsuk for her constructive advice and profound expertise.

\bibliographystyle{apalike}
{\small
\bibliography{iccc}}

\end{document}